# Considerations for developing predictive models of crime and new methods for measuring their accuracy.

Authors: Chaitanya Joshi[1,2], Clayton D'Ath[2], Sophie Curtis-Ham[3] and Dean Searle[4].

## Abstract:

Developing spatio-temporal crime prediction models, and to a lesser extent, developing measures of accuracy and operational efficiency for them, has been an active area of research for almost two decades. Despite calls for rigorous and independent evaluations of model performance, such studies have been few and far between. In this paper, we argue that studies should focus not on finding the one predictive model or the one measure that is the most appropriate at all times, but instead on careful consideration of several factors that affect the choice of the model and the choice of the measure, to find the best measure and the best model for the problem at hand. We argue that because each problem is unique, it is important to develop measures that empower the practitioner with the ability to input the choices and preferences that are most appropriate for the problem at hand. We develop a new measure called the penalized predictive accuracy index (PPAI) which imparts such flexibility. We also propose the use of the expected utility function to combine multiple measures in a way that is appropriate for a given problem in order to assess the models against multiple criteria. We further propose the use of the average logarithmic score (ALS) measure that is appropriate for many crime models and measures accuracy differently than existing measures. These measures can be used alongside existing measures to provide a more comprehensive means of assessing the accuracy and potential utility of spatio-temporal crime prediction models.

## 1. Introduction

Crime prevention is key to reducing crime. One crime prevention strategy is to pre-empt crime and use targeted policing and other measures to prevent it from occurring. The success of this strategy partially rests on how accurately one can predict the time and location of a particular crime before it happens. Such prediction should be possible, in principle, given relevant information (e.g., the location and date/time of past crime, and the socio-demographic or environmental factors that are likely to stimulate crime). However, crime is a dynamic and evolving process complexly related to a wide array of factors, many of which are also dynamic and evolving. While numerous statistical models have been built in the last two decades to predict crime with varying degrees of success, evidence of a rigorous model building process has not necessarily been demonstrated in each instance. Any predictions need to satisfy operational constraints so that they can be acted on. Otherwise, a model which may look good on paper could turn out to be ineffective in practice. As a result, building crime models which predict crime with a great degree of accuracy and are of practical use remains an ongoing research problem (Lee et al. 2020, Santos 2019, Ratcliffe 2019).

Another important issue is how to accurately measure the accuracy of a crime model. Statistical theory contains many standard measures to assess model fit and the predictive accuracy. While some of these measures have been employed in the crime science literature, new predictive accuracy measures have also been developed specifically for crime models. However, some of these measures may have drawbacks and there is, as yet, no consensus on which accuracy measures are the most appropriate for a crime model.

1. New Zealand Institute for Security and Crime Science, University of Waikato, Hamilton, New Zealand.
2. Dept. of Mathematics and Statistics, University of Waikato, Hamilton, New Zealand.
3. Evidence Based Policing Centre, New Zealand Police, Wellington, New Zealand.
4. Waikato District, New Zealand Police, Hamilton, New Zealand.

The aims of this paper are first, to discuss the strengths and limitations of existing accuracy measures and propose new measures and second, to highlight some additional challenges and key considerations involved in developing and assessing predictive crime models. The structure of the paper is as follows. Section 2 reviews existing predictive crime model approaches and measures of accuracy proposed so far in the crime science literature. Section 3 proposes two new measures of accuracy for crime models, namely, the penalized predictive accuracy index (PPAI) and the average logarithmic score (ALS). In Section 4, we propose the use of the expected utility function to combine multiple measures. Section 5 discusses some additional considerations involved in building and testing predictive crime models. Finally in Section 6, we conclude by highlighting our key points and implications for future research and practice.

## 2. Predictive Crime Models and Measures – a review

We first review some of the important crime prediction models developed so far and then the measures of predictive accuracy or operational efficiency that have been used in the crime literature.

### 2.1 A Brief Review of Crime Prediction Models

A large number of different crime models have been developed over time. It is not possible to include all of them in a brief review and more comprehensive reviews exist (e.g., Kounadi et al., 2020). Here, we aim to highlight models that exemplify some of the important types of models that have been proposed.

Early approaches were primarily based on time series analysis and attempted to study how crime rates, as well factors which could influence crime (e.g. unemployment rates, drug use, deterrence, legislative changes) evolved over time, in order to explain the level of crime (e.g. Adams-Fuller, 2001; Cantor & Land, 1985; Corman & Mocan, 2000; Sridharan et al., 2003). Such models have limited utility as predictive models because the causal structure is often weak or partially incorrect (Greenberg, 2001) and because they make community level predictions which may not be as actionable as models which make space and time specific predictions.

The terms *retrospective* and *prospective* have been used to classify the crime models (Johnson, et al., 2007; Caplan, et al., 2011). While such classification has no formal statistical meaning, the terminology is useful to distinguish the models based on the predictive rationale used.

*Retrospective* models use past crime data to predict future crime. These include hotspot based approaches which assume that yesterday's hotspots are also the hotspots for tomorrow. This assumption has empirical justification: research has shown that while hot spots may flare up and cool down over relatively short periods of times, they tend to occur in the same places over time (Spelman, 1995). Hotspot models have typically been spatial models only, not explicitly accounting for temporal variations (e.g. Adams-Fuller, 2001), so seasonal or cyclical patterns could be missed. Retrospective time series models have also been proposed (e.g., Gorr & Olligschlaeger, 2001; Gorr, et al.,2002), and while the more complex of these methods are able to capture various patterns in crime over time, they also increasingly become less user friendly and have to be aggregated to a community level (Groff & La Vigne, 2002), limiting their use for informing patrol patterns.

*Prospective* models use not just past data, but attempt to understand the root causes of crime and build a mathematical relationship between the causes and the levels of crime. Prospective models are based on criminological theories and model the likely prospect of crime based on the underlying causes. It is therefore expected that these models may be more meaningful and provide predictions that are more 'enduring' (Caplan et al., 2011). Prospective models developed so far are based on either socio-economic factors (e.g., *RTM* by Caplan et al., 2011) or the *near*-repeat phenomenon (e.g.,

*Promap* by Johnson et al., 2007; *PredPol* by Mohler et al., 2011). The *near-repeat* refers to the widely observed phenomenon (especially in relation to crimes such as burglary), where a property or the neighbouring properties or places are targeted again shortly after the first crime incident (Johnson, 2007).

Employing a near repeat approach, Johnson et al. (2009) modelled the near repeat phenomenon (i.e., for how far and for how long is there an increased risk of crime) and produced a predictive model named *Promap*. Mohler et al. (2011) modelled the near repeat phenomenon using self-exciting point processes which have earlier been used to predict earthquake aftershocks. This model is available in a software package PredPol. While these two models consider the near repeat phenomenon, they do not consider longer term historical data, taking into account the overarching spatial and temporal patterns. They also do not take into account the socio-demographic factors which can result in crime, and long term changing dynamics in suburbs/communities.

In contrast, *Risk Terrain Modelling* (RTM), developed by Caplan et al. (2011) combines a number of socio-demographic and environmental factors using a regression based model to predict the likelihood of crime in each grid cell. However, this model does not consider historical crime data, thus may not accurately capture the overarching spatial and temporal patterns in crime. It also does not take near repeats into account and thus does not consider short term risks at specific locations. Recent research has demonstrated that RTM can be less accurate than Machine Learning methods that better model the complexity of interactions between input variables, such as Random Forest (Wheeler & Steenbeek, 2020).

Ratcliffe et al. (2016) argue that a model which includes both the short term (near repeat) as well as long term (socio-demographic factors and past crime data) components has a superior 'parsimony and accuracy' compared to models which only include one of those. While this argument is logical, their assertion is based on comparing models using their BIC (Bayesian Information Criterion) values. While BIC is a standard statistical measure to compare models, it measures how well a given model 'fits' or 'explains' the data (e.g., Dobson & Barnett, 2004) and does not directly measure the predictive accuracy (e.g., Shmueli 2010). As discussed in Section 3, BIC is not the most appropriate measure to find the model with the best predictive accuracy. Therefore the assertion made by Ratcliffe et al. (2016) still remains to be verified.

In recent years several attempts have been made to build predictive crime models using artificial neural networks based machine learning algorithms (Rummens et al., 2017; Tumulak & Espinosa, 2017; Wang et al., 2017; Chun et al., 2019). Each of these studies report encouraging results indicating that neural network based models could play an important role in predicting crime in the future. Neural networks are often considered as 'black box' models and a common criticism of such models is that they cannot explain causal relationships. Thus, while a neural network model may be able to predict crime with good accuracy it may not be able to highlight the underlying causal factors and could lack transparency in how it works.

Lee et al. (2020) argued that transparency in exactly how an algorithm works is just as important a criterion as predictive accuracy and operational efficiency. They point out that many of the available crime models are complex, proprietary and lack transparency. They propose a new Excel based algorithm that is fully transparent and editable. It combines the principal of population heterogeneity in the space-crime context with the principal of state dependency (repeat victimization). The authors claim their algorithm outperforms existing crime models on operational efficiency, but not on accuracy. However, they do point to further improvements that could potentially lead to better accuracy.

While individual authors have argued about the strengths of their respective methods, there have been little in terms of independent comparative evaluations. Perry et al. (2013) and Uchida (2014) concluded that statistical techniques used in predictive crime analytics are largely untested and are yet to be evaluated rigorously and independently. Moses and Chan (2016) reviewed the assumptions made while using predictive crime models and issues regarding their evaluations and accountability. They concluded by emphasizing on the need to develop better understanding, testing and governance of predictive crime models. Similarly, Meijer and Wessels (2019) concluded that the current thrust of predictive policing initiatives is based on convincing arguments and anecdotal evidence rather than on systematic empirical research, and call for independent tests to assess the benefits and the drawbacks of predictive policing models. Most recently, Kounadi et al. (2020) conducted a sytematic review of spatial crime models, and concluded that studies often lack a clear reporting of study experiments, feature engineering procedures, and use inconsistent terminology to address similar problems. The findings of a recent randomized experiment (Ratcliffe et al., 2020) suggested that use of predictive policing software can reduce certain types of crime but also highlighted the challenges of estimating and preventing crime in small areas. Collectively, these studies support the need for a robust, comprehensive and independent evaluation of predictive crime models.

## 2.2 Measures for Comparing Crime Models

Being able to identify superior predictive models is central to the goals of predictive through focusing preventive efforts where crime is forecast to occur. Note that a model will be superior subject to a given data and a given set of criteria. It will not be possible to claim superiority over all possible datasets and all possible criteria.

Crime science research has employed both the standard statistical measures as well as measures developed specifically for predictive crime models. There is, however, little consensus as to how to best measure, and compare, the performance of predictive models (Adepeju et al., 2016). Here we provide a brief review of some of these measures. It is not an exhaustive review but highlights measures that exemplify the different types of approaches that have been proposed. Some of these measures assess the ability of predictive models to accurately predict crime (i.e. whether the predictions come true) while others assess their ability to yield operationally efficient patrolling patterns; minimising patrol distance whilst maximising potential prevention gain. Model assessment is typically based on comparing predictions derived from one time period (or a model derived from a 'training' dataset) to observed crimes in a subsequent time period (or an independent 'test' dataset), as distinct from measures of model 'fit' to the original data. To account for variability in predictive accuracy over time, the reported value is often the average value of the measure over several test time periods. Specifically, because we focus on the measures of predictive accuracy or operational efficiency, we do not include the measures of model fit, such as the BIC, in this review.

A natural approach to assess the predictive accuracy of a model is by looking at the distribution of True Positives (TP), False Positives (FP), True Negatives (TN) and False Negatives (FN). Several of the accuracy measures proposed in the crime literature are indeed based on one or more of these quantities. This includes the 'hit rate' (Bowers et al., 2004; Chainey et al., 2008; Johnson et al., 2009; Kennedy et al., 2010; Mohler et al., 2011; Hart & Zandbergen, 2012; Perry et al., 2013; Adepeju et al., 2016; Lee et al., 2017; Rummens et al., 2017) which is the proportion of crimes that were correctly predicted by the model out of the total number of crimes committed in a given time period (TP/(TP+FN)), and is typically applied to hotspots identified by the model. Similarly, a measure termed as 'precision' and defined as the proportion of crimes that were correctly predicted by the model out of the total number of crimes predicted by the model (TP/(TP+FP)) has also been proposed (Brown & Davis, 2006; Rummens et al., 2017). Finally, a measure termed as 'predictive accuracy' (PA) that measures the proportion of crimes correctly classified out of the total number of crime ((TP+TN)/(TP+FP+TN+FN)) has also been used (Mu et al., 2011, Malik et al., 2014, Araujo et al., 2018). Note that while the terms used, namely, hit rate, precision and predictive accuracy may appear to be novel, the measures themselves are well established in statistical literature (e.g., Fawcett (2006)). Thus, hit rate refers to what is also commonly known as the 'sensitivity' of the model, whereas, precision refers to what is also commonly referred to as the 'positive predictive value' or the PPV. Finally, predictive accuracy refers to what is often referred to as 'accuracy' (Fawcett, 2006). A contingency table approach, which takes into account all four – TP, TN, FP and FN – quantities has also been used (Gorr & Olligschlaeger, 2002), and statistical tests of association on such contingency tables have also been applied (Caplan, 2011; Kennedy et al., 2011; Ratcliffe et al., 2016). Rummens et al., 2017 used receiver operating characteristic (ROC) analysis that is fairly common in other domains such as science or medicine (e.g., Brown & Davis, 2006; Fawcett, 2006). ROC analysis plots the hit rate (sensitivity) against the false positive rate 1-specificity (FP/(TN+FP)) at different thresholds.

A related measure that has been developed and widely used in the crime literature is the predictive accuracy index (PAI; Chainey et al., 2008; Levine, 2008; Van Patten et al., 2009; Tompson & Townsley, 2010; Hart & Zandbergen, 2012, 2014; Harrell, 2015; Drawve et al., 2016; Adepeju et al., 2016; Rummens et al., 2017). Hit rate does not take into account the operational efficiency associated with patrolling the hotspot identified. For example, a large hotspot may have a high hit rate simply because it accounts for more crime, yet, such a hotspot will have very little practical value in terms of preventing the crime since it may not be effectively patrolled. PAI overcomes this drawback by scaling the hit rate using the coverage area. If two hotspots have a similar hit rate, the one which has a smaller coverage area will have a higher PAI. Thus, PAI factors in both the predictive accuracy as well as the operational efficiency of the model.

Several other attempts have been made to incorporate operational efficiency into a measure. Bowers et al. (2004) proposed measures such as the *Search Efficiency Rate* (SER) which measures the number of crimes successfully predicted per Km2 and *Area-to-Perimeter-Ratio* (APR) which measures how compact the hotspot is and gives higher scores for more compact hotspots. Hotspots may be compact but if they are evenly dispersed over a wide area then they would still be operationally difficult to patrol compared to if the hotspots were clumped together, for example. The *Clumpiness Index* (CI; Turner, 1989; McGarigal et al., 2012; Adepeju et al., 2016; Lee et al., 2017) attempts to solve this problem by measuring the dispersion of the hotspots. A model which renders hotspots that are clustered together will achieve a higher CI score compared to a model which predicts hotspots that are more dispersed. The *Nearest Neighbour Index* (NNI; Johnson et al., 2009; Levine, 2010) provides an alternative approach to measure dispersion based on the nearest neighbour clustering algorithm.

A model that predicts hotspots that change little over consecutive time periods may be operationally preferred over a model where the predicted hotspots vary more. Measures have been proposed to measure the variation of the hotspots over time. These include, the *Dynamic Variability Index* (DVI; Adepeju et al., 2016) and the *Recapture Rate Index* (RRI; Levine, 2008; Van Patten et al., 2009; Hart & Zandbergen, 2012, 2014; Harrell, 2015; Drawve, 2016; Drawve et al., 2016). One advantage of DVI is that it is straightforward to calculate and does not require specialized software. However, if the actual crime exhibits spatial variation over time then one would expect a good predictive model to capture it and hence the DVI would be higher for that model compared to (say) another model that did not capture this variation, and hence had a lower predictive accuracy. Thus, measures such as DVI need to be considered in conjunction with the predictive accuracy of the model, and not independently. Finally, the *Complementarity* (Caplan et al., 2013; Adepeju et al., 2016) is a visual measure that measures the number of crimes that were uniquely predicted by a given method using a Venn diagram.

Some measures may be arguably superior to others because they account for more aspects of accuracy. For example, PAI could be considered as superior to the hit rate because it also accounts for the corresponding hotspot area. But in other cases, the measures measure accuracy differently. For example, the *hit* rate measures the *sensitivity* whereas the *precision* measures the PPV. Each of them captures the accuracy in some way- but not the other. In such cases, which measure is more appropriate depends on the particular application and the subjective opinions of the analysts. Rather than aiming to find just *one* measure, we recommend using multiple measures to ensure that all aspects of accuracy are assessed. Kounadi et al. (2020) also argued in favor of including complimentary measures. In fact, as we illustrate later in this paper, some measures can be combined in the desired way using the *expected utility* function. The model with the highest expected utility can be considered to be the best model.

The predictive accuracy of a given model will vary over time since the data considered in building the model and the actual number of crimes that happened during the prediction period will vary with time. Therefore, in practice, accuracy obtained over time will have to be somehow summarised. Considering the mean value is important but will not measure the variation in the accuracy observed over time. Therefore, it is also important to consider the standard deviation of the accuracy as well.

A final issue is to test whether the differences in accuracy (however measured) between models are statistically significant. Adepeju et al. (2016) employed the Wilcoxon signed-rank test (WSR) to compare the predictive performance of two different models over a series of time periods. WSR is a non-parametric hypothesis test that can be applied to crime models under the assumption that the difference in the predictive accuracy of the two methods is independent of the underlying crime rate. When comparing multiple models, a correction method such as Bonferroni's has to be applied to ensure that the probability of false positives (in relation to whether the difference is significant or not) is maintained at the desired (usually 5%) level.

## 3. New Measures for Crime Models

Here we explain a possible limitation of the PAI and propose the use of a new measure (Penalised PAI: PPAI) that we have developed to address that limitation. We also propose the use of another measure (Average logarithmic score: ALS) that has recently been used in another domain but that we believe, could be useful for predictive crime models too.

### 3.1 Penalised Predictive Accuracy Index (PPAI)

A limitation of PAI is that as long as the model is correctly identifying a hotspot, it will prefer the model whose hotspot area is smaller simply because of the way PAI is formulated. This drawback is best illustrated by a simple hypothetical example. Suppose an urban area consists of 15 hotspots which together account for 42% of the area but 83% of the crime based on historical data. They have been listed below in decreasing order of their individual PAI (the top hotspots have the highest hit rate and the smallest area). We use *a/A* to denote the proportion of area covered by hotspot/s and *n/N* to denote the proportion of crimes in that hotspot/s, where, *n* denotes the number of crimes in that hotspot and *N* the total number of crimes in the urban area.

| hotspot | 1 | 2 | 3 | 4 | 5 | 6 | 7 | 8 | 9 |
|---|---|---|---|---|---|---|---|---|---|
| a/A | 0.01 | 0.01 | 0.01 | 0.01 | 0.02 | 0.02 | 0.02 | 0.03 | 0.03 |
| n/N | 0.1 | 0.09 | 0.08 | 0.07 | 0.06 | 0.06 | 0.05 | 0.05 | 0.05 |

| hotspot | 10 | 11 | 12 | 13 | 14 | 15 | Total |
|---|---|---|---|---|---|---|---|
| a/A | 0.03 | 0.04 | 0.04 | 0.05 | 0.05 | 0.05 | **0.42** |
| n/N | 0.04 | 0.04 | 0.04 | 0.04 | 0.03 | 0.03 | **0.83** |

Suppose that we are comparing four models: models M-I, M-II, M-III and M-IV. The hotspots identified, their collective relative area, hit rate and PAI are listed in the table below.

| Models | Hotspots identified | a/A | n/N | PAI |
|---|---|---|---|---|
| **M-I** | 1,2,3 | 0.03 | 0.27 | 9.00 |
| **M-II** | 1, | 0.01 | 0.1 | 10.00 |
| **M-III** | 1,5,10,15 | 0.11 | 0.23 | 2.09 |
| **M-IV** | 13,14,15 | 0.15 | 0.1 | 0.67 |

Here, model M-I correctly identifies the top three hotspots (each covering only 1% of the area and together they account for 27% of the crime). Model M-II only identifies the top hotspot and yet, PAI for model M-II is higher than the PAI for model M-I.

PAI (M-I) = $\frac{0.27}{0.03} = 9$ PAI (M-II) = $\frac{0.1}{0.01} = 10$

However, model M-I is clearly superior since it correctly identifies the top 3 hotspots compared to model M-II which only identifies the top one. While smaller hotspot areas are considered relatively easier to patrol, identifying smaller or fewer hotspots, by definition, also means capturing a relatively smaller proportion of crime. As illustrated by above example, PAI could fail to penalize a model which only identifies a subset of the hotspots compared to another model which may be able to correctly capture more hotspots.

To overcome this limitation, we propose a Penalized PAI (PPAI) which penalizes the identification of a total hotspot area that is too small.

$$PPAI = \frac{\frac{n}{N}}{\left(\frac{a}{A}\right)^{\alpha}}$$

The penalization is done by using an extra parameter α whose value is fixed by the user. The value lies between 0 and 1. The value of α is related to the importance of the collective size of the hotspots. In the extreme case when α=0, the size of the hotspots is not important at all. Mathematically, for α =0, the denominator becomes 1 and PPAI reduces to the *hit rate*, i.e. attaching 0 weight to the hotspot area ensures that it is not considered. At the other extreme, α=1 indicates that the hotspot size is extremely important. Mathematically, a unit weight represents no penalty and PPAI reduces to the PAI. Thus, both the *hit rate* and the PAI can be considered as special cases of PPAI. For 0 < α < 1, PPAI would prefer a model that strikes the desired balance between capturing enough hotspots and yet ensuring that the collective hotspot size is not too large for practical considerations.

Next we provide two ways of determining the value of α and two different usages of PPAI.

The first option is to choose α = n/N (hit rate). Choosing α = n/N means that the hotspot area identified by a model is weighed by the relative proportion of crimes (n/N) which take place in that area. Thus, if a model identifies hotspots where fewer crimes take place then it will be penalized more (α smaller) than a model that identifies hotspots where more crimes take place. We illustrate this by calculating PPAI with α = n/N for the hypothetical example considered above.

PPAI (M-I) = $\frac{0.3}{(0.03)^{0.27}} = 0.6958$ PPAI (M-II) = $\frac{0.1}{(0.01)^{0.1}} = 0.1584$

Since model M-II identifies hotspots that account for much smaller proportion of crimes, it is penalized more and as a result has a much smaller PPAI value than model M-I. Suppose we have two additional models to test: models M-III and M-IV. Model M-III identifies the top hotspot and three other smaller hotspots, in total accounting for 23% of the crime (much higher than model M-II), however, these hotspots are also much larger accounting for 11% of the area together. Model M-IV is a clearly an inferior model that identifies the bottom three hotspots. Model M-IV achieves the least score on both PAI as well as PPAI, as one would expect. Model M-III however, achieves a higher PPAI but lower PAI compared to model M-II.

PPAI (M-III) = $\frac{0.23}{(0.11)^{0.23}} = 0.3821$ PPAI (M-IV) = $\frac{0.1}{(0.15)^{0.1}} = 0.1209$

| Hotspot | a/A | n/N | Cumulative a/A | Cumulative n/N | PAI | PPAI (α = 0.9) |
|---|---|---|---|---|---|---|
| 1 | 0.01 | 0.1 | 0.01 | 0.1 | 10.00 | 6.31 |
| 2 | 0.01 | 0.09 | 0.02 | 0.19 | 9.50 | 6.42 |
| 3 | 0.01 | 0.08 | 0.03 | 0.27 | 9.00 | 6.34 |
| 4 | 0.01 | 0.07 | 0.04 | 0.34 | 8.50 | 6.16 |
| 5 | 0.02 | 0.06 | 0.06 | 0.4 | 6.67 | 5.03 |
| 6 | 0.02 | 0.06 | 0.08 | 0.46 | 5.75 | 4.47 |
| 7 | 0.02 | 0.05 | 0.1 | 0.51 | 5.10 | 4.05 |
| 8 | 0.03 | 0.05 | 0.13 | 0.56 | 4.31 | 3.51 |
| 9 | 0.03 | 0.05 | 0.16 | 0.61 | 3.81 | 3.17 |
| 10 | 0.03 | 0.04 | 0.19 | 0.65 | 3.42 | 2.90 |
| 11 | 0.04 | 0.04 | 0.23 | 0.69 | 3.00 | 2.59 |
| 12 | 0.04 | 0.04 | 0.27 | 0.73 | 2.70 | 2.37 |
| 13 | 0.05 | 0.04 | 0.32 | 0.77 | 2.41 | 2.15 |
| 14 | 0.05 | 0.03 | 0.37 | 0.8 | 2.16 | 1.96 |
| 15 | 0.05 | 0.03 | 0.42 | 0.83 | 1.98 | 1.81 |

The second option is to find the optimal value of α so that PPAI will peak at the desired hotspot area. For example, operationally, it may only be possible to effectively patrol, say, 2% of the area at a time. The objective then is to find a model which will identify the best hotspots totaling to 2% of the area. To do this, one first finds the optimal value of α using a grid search algorithm and then uses this value of α to evaluate PPAI for all models under consideration. The model with the highest PPAI is the most suitable for the hotspots totaling 2% of the area in terms of the predictive accuracy. The value of α that is optimal according to this criterion will differ from dataset to dataset.

We illustrate how the grid search method can be used to find the optimal α value using our hypothetical example. First, we list the hotspots in an increasing order of their size (smallest first) and within that using the decreasing hit rate (smallest area with highest hit rate first), as shown in the table above. We then identify the top hotspots that account for 2% of the area. In our case, the top two hotspots together account for the 2% of the area, as desired and together they account for 19% of all crime. The optimal α value is then found by calculating the PPAI for all the cumulative hotspot levels for each value of α from 0.01 to 0.99 and finding the α value that yields the highest PPAI at the desired cumulative level (in this case, 2%). Often, there may be a range of values that satisfy this criterion. For instance, for our example, any value of α between 0.87 and 0.92 will ensure that the PPAI for 2% cumulative level is higher than at any other level (say 1% or 3% or higher). Within this range of values, we find the value for which PPAI for 2% level is the most different from the neighboring levels (1% and 3%). That value of α is the optimal value and in our case it turns out to be 0.9. Using α=0.9 gives us the following PPAI scores for the four models.

PPAI (M-I) = 6.3380 PPAI (M-II) = 6.3096

PPAI (M-III) = 1.6767 PPAI (M-IV) = 0.5515

Since the PPAI was asked to find optimal models for the top 2% hotspot area, the scores are now much different. Model M-I is still the best model, but it is now closely followed by M-II, whereas model M-III which identified hotspots of much larger area has a much smaller PPAI score. Model M-IV is clearly still the worst rated model, as expected. Thus, PPAI parameter α provides flexibility around the importance of hotspot area and even allows to choose α that is optimized for a certain desired hotspot level, constituting an important improvement over PAI.

## 3.2 Average Logarithmic Score (ALS)

In predictive modelling one typically uses two sets of data. A *training* dataset – the data used to fit the model and a *testing* dataset – the data that will be used to test the model predictions. In applications such as crime modelling, where a process evolves over time, the testing and the training datasets typically correspond to data from two distinct time periods. Using a separate testing data ensures that the predictive accuracy is correctly measured. One of the reasons why standard statistical model fitting measures such as AIC, BIC and $R^2$ cannot measure the predictive accuracy of a model very accurately (Shmueli, 2010) is that they use the training data only. This is because their main objective is to measure how well a model explains a given set of data and not how well it can predict unseen or future data.

Average logarithmic score (ALS), first proposed by Good (1952) and later advocated by Gneiting and Raftery (2007) for its technical mathematical properties, computes the average joint probability of observing the testing data under a given model. In simple terms, if the ALS for model A is higher than the ALS for model B then that implies that model A is more likely to produce the testing data than model B. Thus ALS directly measures the predictive accuracy of a model.

ALS = $\frac{1}{N}\sum_{i=1}^{N} \log f(\widetilde{x_i}, \theta)$,

Where, $\tilde{x}$ denotes the testing data, θ denotes the model parameters and f$(\widetilde{x_i}, \theta)$ denotes the probability of observing the testing observation $x_i$ under the model. ALS was used by Zhou et al. (2015) to measure the predictive accuracy of their spatio-temporal point process model to predict Ambulance demand in Toronto, Canada. Similar models have been used for crime (e.g., PredPol, Mohler et al., 2011). ALS is not limited to just one type of models though. It can be used for various types of models including regression based models (e.g.,RTM, Caplan et al., 2011), kernel density estimation (KDE) based models (e.g., Chainey, 2013) and even artificial neural network models as long as the network is designed to predict crime probabilities. Thus, ALS could be a suitable measure for predictive accuracy for a large number of predictive crime models.

Unlike PAI and PPAI, ALS is not restricted to the concept of a hotspot. By definition, ALS considers predictions across the whole region and thus measures the accuracy of the model over the entire region and not just specifically within the hotspots. However, one can restrict ALS to the hotspot region (if so desired) by only considering the testing data that falls within the hotspots in calculating the ALS.

A possible drawback of ALS is that, by definition, it can only be used for models that can compute the probability of crime occurring. For example, if a model simply classifies an area as either a hotspot or not, without being able to quantify the probability of it being a hotspot then ALS cannot be computed for such a model.

## 4. Using Expected Utility to combine multiple measures

As discussed, different measures consider different aspects of predictive accuracy and operational efficiency. While there could be reasons why one of those measures could be considered to be the most appropriate measure for a given analysis, in general, using multiple measures will provide a more complete picture of the performance of a model. It is also possible that multiple measures will rate the given models differently and may not all agree on which model is better.

One way to combine multiple measures while being able to account for the importance of each measure, is by using the concept of expected utility. The notion of utility was first introduced in the context of Game Theory by Neumann and Morgenstern (1947). It is now widely used in Bayesian statistics, game and decision theory and economics (e.g., Robert, 2007; Peterson, 2009; Barber, 2012). More recently, expected utility has been used in *Adversarial Risk Analysis* models that model the actions of the strategic adversary and find the optimal actions for the defender (e.g., Rios et al., 2009; Rios and Rios Insua, 2012; Gill et al., 2016; Joshi et al., 2019).

Finding the expected utility of a model involves considering the utilities (gains or losses) associated each outcome and then taking a weighted sum (expected value), where the weights are proportional to the probabilities of the outcomes to arrive at the net average gain/loss of using the model. The utility can be either positive (indicating that gains outweigh losses) or negative (losses outweigh gains) or 0 (losses = gains). One then chooses the model which has the maximum expected utility.

To illustrate the benefits and usage of expected utility, consider the four measures discussed earlier – the true positives (TP), false positives (FP), true negatives (TN) and false negatives (FN). Each of these measure the predictive performance of a model in their unique way. Taking all four of them in to account will provide a more complete picture of the overall performance of a model. However, this now requires either a multi-dimensional evaluation (since a method with a very high true positives rate, could also have a very high false negative rate, for example) or summarizing the four measures into a single measure in an appropriate way.

One approach is to use the contingency table approach and find the p-value of the Chi-squared statistics to summarize these four measures into one (Gorr & Olligschlaeger, 2002). While such an application of the Chi-squared test will determine if the predictive accuracy of a model was statistically significantly different than a random assignment of hotspots, the Chi-squared test does not test for the direction of the difference nor is it able to be used to identify which of the two models is better. Further, a contingency table approach is unable to weigh the four outcomes differently based on their relative importance. These drawbacks are also applicable to other alternatives for the Chi-squared test, such as a Fisher's exact test. Alternatively, one can use the receiver operating characteristics (ROC) analysis (Rummens et al. 2017), however, this uses just two of the four measures: true positives and false positives.

Applying an expected utility approach, each of the four measures can be weighted and combined. A predictive crime model will label a cell/area as '+' (crime likely to happen) or '-' (crime unlikely to happen). We first find the expected utility of a positive /negative label and then find the expected utility of the model.

Expected utility (+) = % TP × utility (TP) + % FP × utility (FP)

Expected utility (-) = % TN × utility (TN) + % FN × utility (FN)

Expected utility (model) = % '+' x Expected utility (+) + % '-' x Expected utility (-)

In policing, the utilities will be subjective and there are likely no fixed norms about what utility should be attached to which outcome. This is a decision which requires consideration of not only the associated costs and gains of crime and policing responses but also the kind of police response that is considered appropriate for a given community (Fergusson, 2020). These considerations are likely complex and utilities could be different for different crime types, locations and time periods. However, the subjectivity associated with determining utilities will make model assessment more realistic and relevant for each analysis. This feature sets the expected utility measure apart from most other measures discussed.

In illustrating this measure, for the sake of simplicity, we have elected to attach relative utility values between the range +1 and -1, where +1 indicates the most desirable outcome and -1, the least desirable. For example, a correct prediction, whether a TP or a TN, may be considered as highly valuable and therefore utility (TP)= utility (TN) = +1. However, a FP, where the model predicted the cell to be a hotspot but no crime happened during the prediction window, could be considered as a more acceptable and smaller loss (the cost of resources spent) than the FN, where the model predicted the cell to be not a hotspot but a crime did happen, resulting in the cost of that crime to the victim and society (in investigating and dealing with the offence). Therefore, we may assign utility (FP) = -0.5, and utility (FN) = -1.

Suppose we have two models. Model A with a very high TP rate but also a very high FN rate. Model B with a slightly lower TP rate but substantially lower FN rate. Expected utility will enable us to identify which model is better according to the utility criteria. We assume that both the models predict 5% of cells to be hotspots (i.e % '+' = 0.05).

| Model | TP (%) | FP(%) | TN(%) | FN(%) |
|---|---|---|---|---|
| Model A | 85 | 15 | 30 | 70 |
| Model B | 75 | 25 | 45 | 55 |

<u>Model A</u>

Expected utility (+) = 0.85 × 1 + 0.15 × (-0.5) = 0.775

Expected utility (-) = 0.3 × 1 + 0.7 × (-1) = -0.4

**Expected utility (A)** = 0.05 x 0.775 + 0.95 x (-0.4) = **-0.34125**

<u>Model B</u>

Expected utility (+) = 0.75 × 1 + 0.25 × (-0.5) = 0.625

Expected utility (-) = 0.45 × 1 + 0.55 × (-1) = -0.1

**Expected utility (B)** = 0.05 x 0.625 + 0.95 x (-0.1) = **-0.06375**

In the above example, both models have an expected utility that is negative (as a result that 95% of the cells are labeled '-' and the FN rate is high). However, it also reveals that model B has much smaller negative utility despite having a smaller TP rate. Thus, we now know that model B provides better predictions overall after taking account of all the possible outcomes and the utilities associated with each possible outcome.

A limitation of the expected utility is that, it cannot be applied to all accuracy measures. By definition, expected utility can only be applied to measures that correspond to mutually exclusive events. It is not possible to combine say, the PPAI and the ALS using expected utility. In addition to TP, FP, TN and FN, some other related measures too can be combined using expected utility. For example, sensitivity (hit rate) can be combined with specificity (TN/(TN+FP)) and PPV (precision) can be combined with negative predictive value (NPV) which is TN/(FN+TN) and so on.

An additional option is to define a weight measure inspired by the expected utility concept. The probabilities can be replaced by weights that are positive and sum to 1 (just like probabilities) and represent the relative importance of each measure. We illustrate this method with the same hypothetical example, where we have the two models A and B, but we now wanted to use two different measures, namely hit rate and precision. We can combine them by taking the weighted sum of their scores. Using the hypothetical TP, FP, TN and FN values already considered, we can calculate the hit rate and precision for these two models as shown below. Here, Model A is the better model according to precision but Model B is the better model according to the hit rate.

| Model | Hit rate | Precision |
|---|---|---|
| Model A | 0.06 | 0.85 |
| Model B | 0.067 | 0.75 |

Note that since we have assumed %'+' =0.05' for both the models, hit rate (A) = 85*0.05/(85*0.05+70*0.95) = 0.06 and similarly, hit rate (B) = 75*0.05/(75*0.05+55*0.95) = 0.067. Also note that in this case because each model yields hotspots of the same area (5%), the PAI ranking will be exactly same as the hit rate ranking. An analyst might want to give a considerably high weight to the hit rate (say w =0.7) and the remaining (1-w =0.3) to the precision. The weighted aggregate for the two models then become:

Model A: 0.7*0.06 + 0.3*0.85 = 0.297

Model B: 0.7*0.067 + 0.3*0.75 = 0.271

Taking into consideration the relative importance of the two measures, we can now see that Model A is the better model overall. That this ranking does not match the ranking obtained using expected utility on TP, FP, TN and FN is not surprising because different measures were used here and combined differently.

Another challenge is that when combining different measures one needs to ensure that the results are not distorted due to scaling. This issue did not arise when combining hit rate and precision, since both will always take values between 0 and 1. However, other measures could yield high positive values or even large negative values (in case of ALS, being aggregation of log probabilities), simply because of the way they are defined. To avoid distortion due to scaling, weighted aggregation should only be performed after standardizing the scores (0 mean and unit standard deviation), where possible. Alternatively, models could be ranked according to each measure and the weighted aggregation can be performed on the ranks.

## 5. Additional Considerations when Choosing and Comparing Models

Development and comparison of predictive models is not straight-forward; it requires careful consideration of various issues of both a technical and practical nature. Further to our proposed methods to support robust and comprehensive evaluation of crime prediction models, here we elaborate on some additional issues to consider in selecting and assessing models.

## Technical considerations

Performance of a predictive model depends on several key technical factors. These include the type of model used, the variables included in the model, the tuning or calibration of any free parameters, the sparsity of the data, the predictive window used and finally, the way in which its accuracy is measured.

From a purely technical point of view, a model should only be considered if it is appropriate given the aims of the analysis and if any founding mathematical assumptions made by the model have been satisfied by the data. Additionally, one must consider the practicalities of data pre-processing or data-coding needed, the ease or complexity of implementing the models and the computational and financial costs needed to run these models.

During the model building process, one or more free parameters may need to be assigned a value. Common examples include the bandwidth in a kernel density estimation model or a smoothing parameter in a time-series model. A common approach with crime data is to aggregate it over rectangular grid cells and then use this aggregated data for analysis. Here, the cell size is also a free parameter. Predictions will likely be sensitive to the values assigned to such free parameters (Chainey, 2013; Adepeju etal, 2016). Their values must thus be assigned based on some theoretical or empirical justification and not arbitrarily. It is also advisable to perform some robustness analysis to understand the sensitivity of the predictions to the assigned values.

The quality and usability of the predictions are usually a product of the type of the model used and the factors included. For example, a model which considers socio-demographic factors but not spatial or temporal factors may yield predictions which are independent of space and time. Therefore, it may be of lesser operational value than a model which also considers space and time and offers predictions specific to a given location and time. However, such a model is also likely to require a larger and denser dataset in order to provide accurate predictions. Further, the predictions of a model are usually only valid for the range of the predictor variables considered in the analysis. These are not likely to be valid or accurate for values of the predictor variables outside that range as well as time periods too far out in the future. The factors used may change with space and over time; the model assumptions may not be valid in the future or to a different area.

The size of the dataset and the density or sparsity of the data are also critical factors. Size relates not only to the number of observations, but also to the amount of information available for each observation and its spread across space and time. Typically, larger data contains more information and allows for more advanced models incorporating more variables. Its density or sparsity depends on how many observations are available with each level of a given factor. For example, in a small town with low crime rate, it may only be reasonable to include either the spatial factor or the temporal factor but not both. Implementing a model on data that is too small or sparse may lead to overfitting and hence, poor predictive accuracy. Typically, data become sparser as more factors are considered affecting the optimal cell size and other free parameters.

Finally, multiple measures are available to measure the performance of a model and compare multiple models. Some measures consider the goodness of fit, some others look at the predictive accuracy and indeed some others may take into account the operational utility. The same model may fare differently when assessed using different measures. Therefore, one needs to identify the most appropriate measure for the study and then use that measure to identify the best performing model. For example, as discussed earlier, traditional goodness of fit measures such as AIC (Akaike Information Criterion) or BIC (Bayesian Information Criterion) may be more appropriate when comparing models to see which one of them 'explains' the data better. However, as noted above, these do not measure predictive accuracy and are therefore not the most appropriate measures for finding the model that has the highest predictive accuracy. Another, more complex issue is how to identify the best model when different measures point to different models. We have proposed one solution, considering the expected utility measure or the weighted aggregate measure discussed in Section 4. Either of these options involves making a subjective choice, meaning that someone with a different set of preferences for measures could choose a different model as the best model for the same data and using the same set of measures. Further, different analysts may prefer to use other measures altogether, hence are likely to arrive at different conclusions about relative model performance. Analysts should recognize this subjectivity and provide a clear rationale for the choice of models, performance measures and weights used to combine the measures. This will go towards providing more transparency and reproducibility of studies of crime prediction models.

### Other considerations

Although the focus of this manuscript is on technical considerations, there are several practical issues to consider in deciding on the most suitable predictive crime model including ethical and legal aspects. As discussed earlier, Lee et al. (2020) have argued that transparency in exactly how an algorithm works is just as important a criterion as predictive accuracy and operational efficiency. But it is also important that the data used is obtained using best practices, is accurate and not a product of racially biased or unlawful practices. Recent research (Richardson et al., 2019) has highlighted the ramifications of using predictive policing tools informed by 'dirty data' (data obtained during documented periods of flawed, racially biased and sometimes unlawful practices and policies). Predictive policing models using such data could not only lead to flawed predictions but also increase the risk of perpetuating additional harm via feedback loops.

Finally, the choice of model is also a choice of policing theory. As Ferguson 2020 argues, when purchasing a particular predictive technology, police are not simply choosing the most sophisticated predictive model; the choice reflects a decision about the type of policing response that makes sense in their community. Foundational questions about whether we want police officers to be agents of social control, civic problem-solvers, or community partners lie at the heart of any choice of which predictive technology might work best for any given jurisdiction.

## 6. Discussion and Summary

Significant research effort has focused on developing predictive crime models. Many different models have been proposed and claims made about the superiority of a given model. As pointed out by several literature reviews, these claims have often not been based on rigorous, independent and impartial assessment. While finding or developing appropriate measures to assess and compare the performance of these models has not received nearly equal attention, a few measures have been developed specifically for crime application. However, existing measures have limitations and further work on finding appropriate measures is needed. The ALS measure described above is one such measure. It has been used in other domains, measures aspects not measured by existing measures, and could be used for a wide variety of crime models.

It is worth emphasizing that there can be no single model or indeed no single measure that is superior over all others at all times. Future studies should explicitly explain why certain measures were considered the most appropriate for the problem at hand and demonstrate how a given model performs according to those measures. As discussed, the accuracy achieved not only depends on the model but on the quality of the data and its density or sparsity. The quality of the data refers not only to omissions or inaccuracies in the data but also on whether the data reflects flawed or biased practices. The choice of model is not only about the predictive accuracy or the operational efficiency possible, but also about the choice of policing theory and what is appropriate for the community in question.

It is therefore important to develop measures which empower the practitioner with a certain level of flexibility to tweak the measure so that it is the most appropriate for a given situation. The penalty parameter in the PPAI measure that we propose does precisely that. It empowers the practitioner to choose the right balance between capturing enough crime and operational efficiency for the problem at hand. Since different measures measure different aspects of accuracy and efficiency, it may be desirable to use multiple measures to assess models. However, this can be challenging because different measures could rank models differently and combining the measures may not be straightforward. A possible solution is a mathematical function that empowers the user with the flexibility to combine multiple measures in a way that is desirable for the problem at hand. We suggest using the expected utility function and its extension, the weighted aggregate for this purpose.

The concept of utility or weights reflects subjective inputs and therefore could be perceived as undesirable. However, they in fact model the human decision-making process. Decision makers have different set or preferences and value systems which are often reflected in the decisions they make. The use of weights enables the practitioner to translate their thought process in an objective mathematical equation and ensures that the equation will identify the correct model once the weights have been elicited according to the user's preferences. Since every dataset, prediction problem and context is unique, a seemingly 'objective' or 'one size fits all' solution is unlikely to be the right approach. Instead, we advocate solutions that empower practitioners to tailor their assessment to their particular problem and to clearly document their decision-making. In doing so, future studies are more likely achieve standards of transparency, reproducibility and independence that will move crime prediction forward as a science.